\documentclass[10pt,conf]{IEEEtran}
\ifCLASSINFOpdf \else \fi
\usepackage{geometry}
\usepackage{multirow}
\usepackage{bbold}
\usepackage{ifpdf}
\usepackage[normalem]{ulem} 
\usepackage{subcaption}%
\usepackage[cmex10]{amsmath}
\usepackage{amssymb}
\usepackage{verbatim}
\usepackage{algorithm}
\usepackage{algorithmic}
\usepackage{array}
\usepackage{latexsym}
\usepackage{color}
\usepackage{textgreek}
\usepackage{subscript}
\usepackage{rotating}
\usepackage{booktabs,multirow}
\usepackage{mdframed}	
\usepackage{tabularx}
\usepackage{amsmath}
\usepackage{makecell}
\usepackage{tabto}
\usepackage{mathrsfs}
\usepackage{url}
\usepackage{cite}
\usepackage{listings}
\usepackage{array}
\usepackage[table]{xcolor}
\usepackage{ragged2e}
\usepackage{booktabs}

\usepackage{graphicx}
\usepackage{comment}
\usepackage[cmex10]{amsmath}
\usepackage{amssymb}
\usepackage{bbold}
\usepackage{float}
\usepackage[cmex10]{amsmath}
\usepackage{amssymb}
\usepackage{verbatim}
\usepackage{array}
\usepackage{latexsym}
\usepackage{color}
\usepackage{tabularx} 
\usepackage{textgreek}
\usepackage{subscript}
\usepackage{rotating}
\usepackage{booktabs,multirow}
\usepackage{mdframed}	
\usepackage{tabularx}
\usepackage{amsmath}
\usepackage{makecell}
\usepackage{tabto}
\usepackage{mathrsfs}
\usepackage{url}
\usepackage{cite}
\usepackage{listings}
\usepackage{array}
\usepackage[table]{xcolor}

\definecolor{pinkpurple}{rgb}{0.6, 0.1, 0.9} 
\usepackage{algorithmic}
\usepackage{array}
\usepackage{soul}
\sethlcolor{green}
\usepackage{url}
\usepackage{pifont}
\usepackage{xcolor}
\usepackage{multirow}
\usepackage{graphicx} 

\begin{document}


\title{Partitioned Task Offloading for Low-Latency and Reliable Task Completion in 5G MEC}

\author{
	Parisa~Fard~Moshiri,~Murat~Simsek~\IEEEmembership{Senior~Member,~IEEE},~and~Burak Kantarci,~\IEEEmembership{Senior Member,~IEEE}
\thanks{
The authors are with the School of Electrical Engineering and Computer Science at the University of Ottawa, Ottawa, ON, K1N 6N5, Canada.
E-mail: \{parisa.fard.moshiri, murat.simsek,burak.kantarci\}@uottawa.ca}  
}

\maketitle
\thispagestyle{empty}
\pagestyle{empty}
\begin{abstract}
The demand for MEC has increased with the rise of data-intensive applications and 5G networks, while conventional cloud models struggle to satisfy low-latency requirements. While task offloading is crucial for minimizing latency on resource-constrained User Equipment (UE), fully offloading of all tasks to MEC servers may result in overload and possible task drops. Overlooking the effect of number of dropped tasks can significantly undermine system efficiency, as each dropped task results in unfulfilled service demands and reduced reliability, directly impacting user experience and overall network performance. In this paper, we employ task partitioning, enabling partitions of task to be processed locally while assigning the rest to MEC, thus balancing the load and ensuring no task drops. This methodology enhances efficiency via Mixed Integer Linear Programming (MILP) and Cuckoo Search, resulting in effective task assignment and minimum latency. Moreover, we ensure each user's Resource Blocks allocation stays within the maximum limit while keeping latency low. Experimental results indicate that this strategy surpasses both full offloading and full local processing, providing significant improvements in latency and task completion rates across diverse number of users. In our scenario, MILP task partitioning results in  24\% reduction in latency compared to MILP task offloading for the maximum number of users, whereas Cuckoo search task partitioning yields 18\% latency reduction in comparison with Cuckoo search task offloading.

\end{abstract}


%
\IEEEpeerreviewmaketitle

\section{Introduction}

The necessity for Multi-Access Edge Computing (MEC) arises from the increasing demand for real-time data processing, driven by the increasing number of User Equipment (UE), creating large amounts of data at the network's edge. Traditional cloud models face difficulties with this data load, given the latency and bandwidth limitations \cite{dong2024}. MEC addresses these challenges by relocating processing closer to end users, hence guaranteeing low-latency \cite{zhang2024,ferrag.comst.2023}.
The implementation of 5G networks has increased the demand for MEC, as edge computing utilizes 5G's potential for ultra-low latency and high bandwidth \cite{butt2024}. The growing volume of data-intensive applications exerts greater pressure on network bandwidth, which MEC reduces by processing data close to its source, reducing the need to transfer large amounts of raw data over long distances. This enhances bandwidth efficiency by transmitting only the most critical data, such as aggregated findings or processed data, instead of all the raw data produced by UEs \cite{akhlaqi2023}.

In MEC, task offloading is a crucial procedure that transfers computational tasks from user devices to edge servers, significantly assisting resource-limited devices. Offloading algorithms are significantly impacted by the computational capabilities of mobile devices and the resources available at the MEC, ensuring a proper balance between performance and resource utilization \cite{yumeng2023}. Nevertheless, offloading introduces overheads. For instance, when multiple tasks are assigned to MEC servers, the servers' limited processing capacity can extend execution time. To address this, intelligent offloading decision strategies are required to improve performance. By intelligent task offloading, latency can be minimized; however, this can result in dropped tasks due to missing deadlines in MEC servers \cite{moshiri2024interplay}. 
A crucial method for improving task offloading is task partitioning, which entails distributing tasks between local processing and MEC offloading \cite{patsias2023}. This method involves processing certain components of the task locally on the UE, while assigning the rest to the MEC server. By processing some portion of tasks locally, the risk of overloading the MEC servers is reduced and with additional constraints we can ensure that tasks aren't dropped.

A thorough review of existing literature reveals that the optimization of combined dropped task and latency minimization has not been extensively explored, particularly focusing on the impact of task partitioning for minimizing latency while ensuring no tasks are dropped. In our previous work \cite{moshiri2024}, we focused on offloading strategies to minimize latency and reduce the dropped task ratio, without leveraging local resources or employing task partitioning between local devices and MEC servers.

Building on that foundation, this study utilizes task partitioning, a significant enhancement that eliminates dropped tasks by efficiently dividing each task between local processing and MEC servers. This parallel processing approach ensures that all tasks are fully processed while further optimizing performance.
Furthermore, efficient resource allocation within the 5G-enabled MEC infrastructure further optimizes task processing by managing resource blocks (RB), ensuring that MEC servers handle substantial computational loads efficiently and the maximum available bandwidth is used efficiently. We incorporate resource blocks in our optimization problem, a feature not considered in our previous work \cite{moshiri2024}. Each user’s allocation of resource blocks is designed to remain within the maximum number of RBs available, optimizing task distribution to prevent network congestion and maintain low latency.
Based on our previous research \cite{moshiri2024}, MILP outperformed other optimization techniques we applied. Thus, we have integrated it here, along with Cuckoo search, to achieve our optimization goals.

The main contribution of this paper is as follows:
\begin{enumerate}
    \item We formulated a comprehensive optimization framework for task partitioning that guarantees zero task drops and enables a comparative analysis of local processing, full offloading, and partitioning approaches across varying user scales. Additionally, we ensure that each user’s allocation of RBs remain within the available maximum number of RBs, while maintaining low latency.
    \item We evaluated our approach using Cuckoo Search and conducted a performance comparison with MILP, demonstrating the effectiveness of our optimization method.

\end{enumerate}

The literature review is located in Section II. The system model is covered in Section III. Performance is discussed and evaluated in Section IV. Section V concludes.

\section{Related Work}

Task partitioning and offloading in MEC is a developing field focused on optimizing the distribution of computational tasks between local devices and edge servers.  Some researchers concentrate on static offloading strategies in which complete tasks are transferred to the edge server, which thereafter manage the processing. Certain methodologies tackle this issue by modeling it as an optimization problem, including Genetic Algorithm (GA) \cite{moshiri2024} and mixed integer nonlinear programming (MINLP) \cite{zheng2021}, and so on. According to the particular issue and scenario, the most effective optimization algorithm is selected to guarantee optimal outcomes. While assigning all tasks to servers reduce the computational burden on mobile devices, it results in inefficiencies as tasks frequently experience delays or are not processed in time, leading to high latency and dropped tasks.

To address these issues, partial offloading emerged as a solution, leveraging parallelism between local and remote executions. It enables the splitting of tasks into parts, with certain tasks executed locally and others assigned to servers. Yi et al. \cite{yi2022} aim to minimize the total latency of a multi-server MEC system in executing all user-offloaded tasks. The problem is formulated as a MINLP problem that simultaneously addresses user-server association,  partitioning, and resource allocation. Xia et al. \cite{xia2023} formulate the issue as a min-max optimization problem, aiming to minimize the maximum delay between task offloading and computation.  Additionally, Multi-Temporal Slot Offloading MTSO and Multi-Cell Multi-Time Slot Offloading (MCMTSO) algorithms are introduced to improve the optimization of subtasks offloaded in each time slot. However, the aforementioned strategies don't guarantee that all tasks would be processed, especially in dynamic network conditions where task drop rates could still be significant. 

Advanced methodologies investigate the application of artificial intelligence and Reinforcement Learning (RL) to enhance task offloading efficiency. Wang et al. \cite{wang2023} develop a collaborative dual-agent deep RL technique to jointly optimize task partitioning and virtual network function (VNF) placement in MEC. This strategy aims to minimize latency by training agents to adjust to task requirements. In another study by Gao et al. \cite{gao2023} a layer-level partitioning strategy for Deep Neural Network (DNN) tasks is proposed. Aggregative game theory is used to improve the placement and processing of DNN layers among devices and servers.

Despite the potential of DNNs in improving task partitioning, these methods frequently necessitate large-scale training data and are computationally expensive, presenting difficulties for resource-limited environment.  Although, RL is exceptionally adaptive, it has significant computational costs, long convergence periods, and the potential to become stuck in sub-optimal solutions during exploration. Conversely, optimization approaches yield consistent solutions, effectively identifying near-optimal and/or optimal results.

Comprehensive analysis of literature indicates that the joint optimization of minimizing dropped tasks and latency has not been adequately explored, particularly concerning the effects of task partitioning on latency reduction while ensuring  no task drops. In our prior researches \cite{moshiri2024interplay}, we investigated offloading mechanisms aimed at minimizing latency and decreasing the drop task ratio. This study extends the methodology by incorporating task partitioning, guaranteeing the proper processing of all tasks through the parallel utilization of local resources and MEC servers and ensuring 0\% dropped task ratio

\begin{figure*}[!hbt]
        \centering
        \includegraphics[width = 0.55\textwidth, trim=1.2cm 3cm 1.2cm 1.2cm,clip]{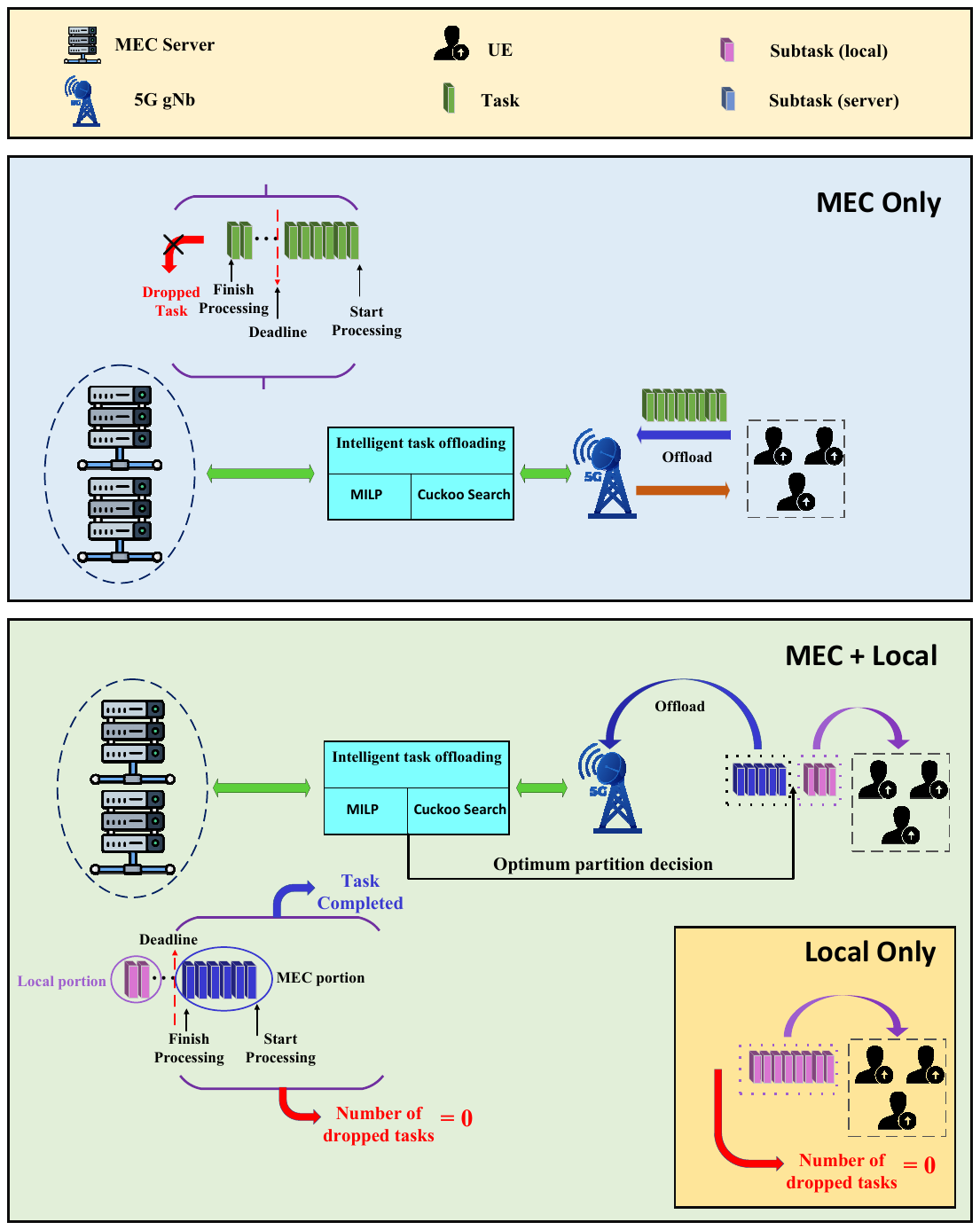}
        \caption{ General concept }
        \label{fig:scenario} 
\end{figure*}

 \begin{table}[!hbt]
\centering
\renewcommand{\arraystretch}{1.4} 

\caption{Notation table for parameters used in the optimization problem.}
\label{tab:notation}
\begin{tabularx}
{0.4\textwidth}{|c|X|}
\hline
\textbf{Parameter} & \textbf{Description} \\
\hline
\( N \) & Set of tasks \\
\hline
\( M \) & Set of servers \\
\hline
\( B_i \) & Bandwidth of channel \\
\hline
\( p_0 \) & Transmission power of UE \\
\hline
\( g \) & Channel gain \\
\hline
\( n \) & Density of noise power \\
\hline
\( S_i \) & Size of the task\\
\hline
\( r_i \) & datarate for UE of task i\\
\hline
\( t_{d_i} \) & Deadline of task \( i \) \\
\hline
\( t_{a_i} \) & Arrival time of task \( i \) \\
\hline
\( t_{s_i} \) & Start  processing time of task \( i \) \\
\hline
\( T^{\text{p}}_{i} \) & processing time of task \( i \) on MEC server\\
\hline
\(T^{\text{lp}}_{i}\) & processing time of task \( i \) on local user\\
\hline
\(T_{i}^c\) & the computational latency for task \(i\) \\
\hline
\(T_{i}^t\) & the communication latency for task \(i\) \\
\hline
\( T^{\text{w}}_{i} \) & Waiting time for task \( i \) \\
\hline
\( RB_{i} \) & Number of resource blocks assigned to task i\\
\hline
\(x_{ij} \) & Binary decision variable which is 1 when task i is assigned to server j.\\
\hline
\(L_{\text{local}}\) & Total latency for local computing (UE) \\
\hline
\(L_{\text{offloading}}\) & Total latency for offloading to MEC server \\
\hline
\(D\) & Number of dropped tasks \\
\hline
\(p_i\) & portion of task \(i\) that is processed locally\\
\hline

\end{tabularx}

\end{table} 
\section{System Model}

\subsection{Task Offloading and Partitioning}
\figurename\hspace{0.1pt}\ref{fig:scenario}  demonstrates how task offloading operates in a 5G-MEC environment. We compare three scenarios: MEC only (full offloading), Local only (Full local), and MEC+Local (partitioning). The main goal of "intelligent task offloading" is to provide low latency and number of dropped tasks.
UEs transfer their tasks to MEC servers within the 5G network, utilizing the servers' stronger processing capabilities. The system seeks to ensure that tasks are completed within specified deadlines, categorizing any tasks that exceed the deadlines as dropped.  To ensure we have no dropped tasks and minimum latency, we utilize task partitioning where system partitions each task into smaller subtasks, distributing them between local and MEC server. Some portions are processed locally, while the rest are offloaded to the MEC servers. The optimal distribution of subtasks between local and server processing is determined through an optimization approach aimed at minimizing latency. In the last scenario, when we only use local only (Full local), we don't have any dropped tasks but we can have bigger latency, due to limited processing capability of UEs. By strategically balancing processing loads between local devices and MEC servers, the system ensures efficient task handling, minimizing total latency, and ensuring zero dropped tasks.

\subsection{The optimization Problem}
The optimization problem for offloading and partitioning includes communication and computation latency. All of the required notation for mathematical formulas is briefly explained in \tablename \hspace{0.1pt} \ref{tab:notation}. We  use  a binary decision variable, \( x_{ij} \), for assigning tasks to servers \cite{moshiri2024} where:

\begin{equation}
x_{ij} = 
  \begin{cases} 
    1 & \text{ task } i \text{ is assigned to server } j  \\
    0 & \text{otherwise}
  \end{cases}
  \label{eq:two}
\end{equation}

We formulate the waiting time for task \(i\), \(T^{\text{w}}_{i}\), as in (\ref{eq:compwaiting}), according to start processing time (\(t_{sp_i}\)), and arrival time (\( t_{a_i} \)). This waiting time represents the delay experienced by the task before it begins execution at the MEC server. The total computational latency accounts for the time taken from the moment the task arrives at the system until it completes processing, including both the waiting time and the actual processing time. Thus, the total computational latency, \(T_{i}^c\), is formulated as (\ref{eq:comp}). The processing time for task \(i\), denoted as \(T_{i}^p\), is obtained from \cite{wang2021}.

\begin{equation}
T^{\text{w}}_{i} = t_{\text{sp}_i} - t_{a_i}
\label{eq:compwaiting}
\end{equation}

\begin{equation}
T_{i}^c = \sum\nolimits_{i=1}^N (T_{i}^p + T_{i}^w)
\label{eq:comp}
\end{equation}

The roundtrip communication latency can be formulated as  in  (\ref{eq:comlatency}): 
\begin{equation}
T_{i}^t = \sum_{i=1}^N 2 \times \frac{S_i}{r_i}
\label{eq:comlatency}
\end{equation}

\begin{equation}
r_{i} = B_{i} \log_2 \left(1 + \frac{p_0 g}{n}\right)
\label{eq:datarate}
\end{equation}
Where \(S_i\) is size of the task \(i\) and \(r_{i}\) is the datarate for the UE of task \(i\), which is calculated as (\ref{eq:datarate}). In (\ref{eq:datarate}), \(B_i\) denotes the bandwidth of the channel for the UE of task \(i\), \(p_0\) is the transmission power of UEs which is identical for all, \(g\) represents the channel gain, and \(n\) is the density of the noise power of the channel  \cite{elgendy20}.

The general objective function is formulated in (\ref{eq:obj1}) and (\ref{eq:partition}):
\begin{equation}
\min \sum_{j=1}^{M} \sum_{i=1}^{N} \left( p_i \times L_{\text{local}} + (1-p_i) \times (L_{\text{offloading}} \times x_{ij} + D \right))
\label{eq:obj1}
\end{equation}

\begin{equation}
\min  \sum_{j=1}^{M} \sum_{i=1}^{N} \left(
p_i \times T^{\text{lp}}_{i} + (1-p_i) \times \left( (T^{\text{c}}_{i} + T^{\text{t}}_{i}) \times x_{ij} + D \right) \right)
\label{eq:partition}
\end{equation}

Where \(T^{\text{lp}}_{i}\) is the processing time for local unit, \(p\) is the portion of a task that is processed locally and \(D\) is number of dropped tasks, which is derived from the complement of the sum of \(x_{ij}\) over all servers \(j\) for each task \(i\) \cite{moshiri2024interplay}. For offloading, we minimize latency and number of dropped tasks, while for partitioning, we minimize latency and ensure no task is dropped. Accordingly, \(p\) varies for different scenarios as below:

\begin{itemize}
    \item $p_i = 1$   : Local only
    \item $p_i = 0$ : Offload only
    \item $0 \leq p_i \leq 1$ : Partition
    \label{eq:p}
\end{itemize}

\textbf{Constraints}: In (\ref{eq:tsp1}), the first task will begin processing upon arrival. In (\ref{eq:tsp}), the start processing time of a task is defined as the period between its arrival time and task deadline, including processing and communication latency. It ensures that the task begins processing at a time that permits it to be completed before the deadline, taking into account communication latency. The constraint in (\ref{eq:c1}) states that each task can only be allocated to one server. In (\ref{eq:rb}), the resource blocks assigned to UE of task \(i\), \(RB_i\), are limited by the maximum number of available resource blocks \(RB_{max}\). 
The linearization for this problem is done according to \cite{moshiri2024}.

\begin{equation}
    S.t. : 
    t_{\text{s}_1} = t_{a_1}
    \label{eq:tsp1}
\end{equation}
\begin{equation}
    t_{a_i} \leq t_{\text{s}_i} \leq t_{d_i} - T_{i}^p - T_{i}^t
    \label{eq:tsp}
\end{equation}
\begin{equation}
\sum\nolimits_{j=1}^M x_{ij} \leq 1   \quad \forall i
\label{eq:c1}
\end{equation}

\begin{equation}
    RB_i \leq RB_{\text{max}}
    \label{eq:rb}
\end{equation}

For partitioning, we introduce an additional constraint (\ref{eq:d}) to ensure that tasks aren't dropped. Additionally, for partitioning, each \(i\) in the constraints (\ref{eq:tsp1}) to (\ref{eq:rb}) represents the portion (subtask) of task \(i\) that is offloaded.

\begin{equation}
    D = 0
    \label{eq:d}
\end{equation}

\begin{table}[!hbt]
\centering
\renewcommand{\arraystretch}{1} 
\caption{Parameter values in simulations} 
\begin{tabular}{|c|c|}
\hline
\textbf{Parameters} & \textbf{Value}   \\   \hline
Number of UEs & 50,100,200,400   \\   \hline
Max number of tasks & 400   \\   \hline
Number of servers & 2   \\   \hline
Number of CPU in each server & 1   \\   \hline
Max bandwidth & 20Mhz   \\   \hline
noise power & 100dbm   \\   \hline
transmit Power & 200 mW \\   \hline
\end{tabular}
\label{tab:parameters} 
\end{table}
\section{Performance Analysis}
The simulations are conducted using a computer equipped with a Core i7 CPU, 4080Ti GPU, and 32GB of RAM. The parameters are set according to \cite{moshiri2024}, as shown in Table~\ref{tab:parameters}. With a 20 MHz bandwidth (and after subtracting the 10\% guard band, resulting in an effective bandwidth of 18 MHz), the max number of RBs is 100. We simulate the 5G gNB with two MEC servers, each equipped with one CPU for processing tasks. The tasks follow a Poisson arrival process and involve different sizes of images \cite{wang2021}. The processing time were obtained from \cite{wang2021} for both local and MEC.  For offloading, all tasks are directed to the MEC servers with the aim of minimizing latency and reducing the number of dropped tasks. In task partitioning, each task is divided into subtasks, and the objective is to determine the optimal portion for local and server processing to minimize latency and ensure no task drops, according to the objective function defined. We compare the performance of MILP with Cuckoo Search, as MILP has shown superior results in our previous work \cite{moshiri2024interplay}. The Cuckoo Search algorithm is configured with 25 nests as potential solutions within the search domain. It runs for 100 iterations, allowing it to thoroughly explore and optimize the solution space. The abandonment probability is set to 0.25, promoting exploration by discarding less optimal solutions to enhance the chances of finding better alternatives. The Lévy flight exponent (Lambda) is set to 1.5, determining the step size, which balances local and global exploration for efficient navigation through the solution space. These parameters have been carefully selected after extensive testing and fine-tuning.


 Cuckoo search as a metaheuristic algorithm navigates the solution space through a combination of random search and Lévy flights. Its ability to balance exploration and exploitation makes it especially efficient for large-scale, complicated issues such as task offloading, where the solution space may be extensive and dynamic. Conversely, the precision of MILP derives from its mathematical formulation, ensuring that each step of the solution process is based on logical and algebraic rules, resulting in a precise and optimal solution. Comparing MILP with Cuckoo Search is beneficial, since MILP employs a precise methodology to identify optimal solutions, whereas Cuckoo Search provides an adaptive, nature-inspired strategy that adeptly explores various  solution spaces. This comparison provides insights into the performance of a precise optimization approach, such as MILP, compared to an adaptable algorithm like Cuckoo Search. This analysis underscores the efficiency of MILP, while also demonstrating Cuckoo Search's potential as an alternative for optimizing task offloading and partitioning.

\begin{figure}[!t]
 \centering
    \begin{subfigure}{0.3\textwidth}
        \centering
        \includegraphics[width =\textwidth, trim= 0cm 0cm 0cm 0cm ,clip]{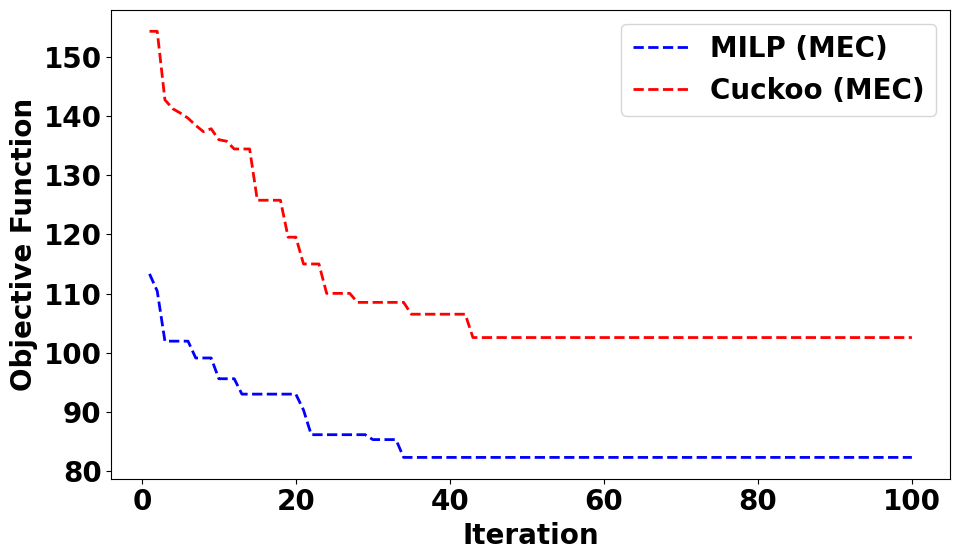}
        \caption{  MEC only}
        \label{fig:convergence1}
    \end{subfigure}
    \begin{subfigure}{0.3\textwidth}
        \centering
        \includegraphics[width =\textwidth, trim= 0cm 0cm 0cm 0cm ,clip]{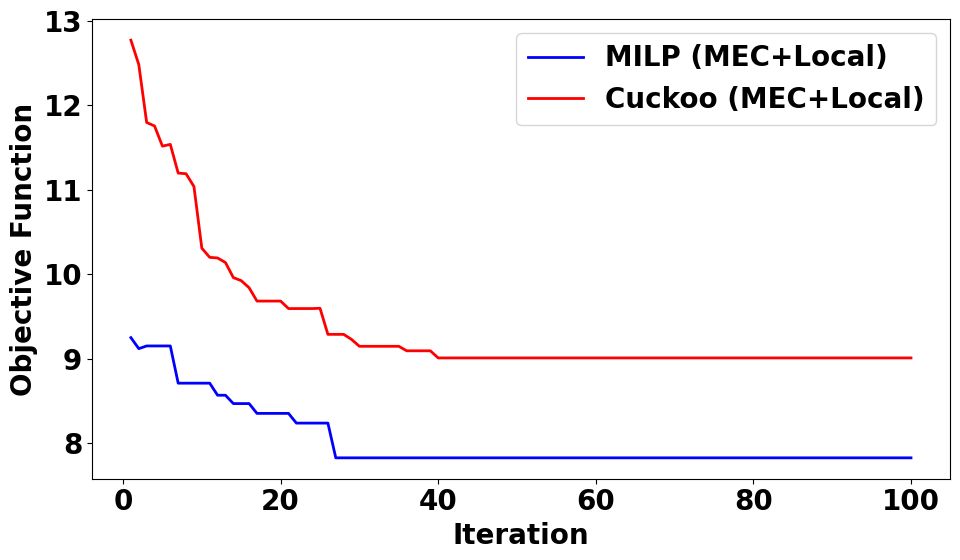}
        \caption{ MEC+Local}
        \label{fig:convergence2}
    \end{subfigure}
    \caption{ Average objective function value for 400 users}
    \label{fig:convergence}
\end{figure}

The convergence plots for MILP and Cuckoo Search in both scenarios is plotted in  \figurename\hspace{0.1pt}\ref{fig:convergence}, which provide insights into their performance, stability, and optimization quality. Since the objective function values for offloading (MEC only) is much bigger than partitioning (MEC+Local), due to considering number of dropped tasks, we plotted them separately. In both cases, \figurename\hspace{0.1pt}\ref{fig:convergence1} and \figurename\hspace{0.1pt}\ref{fig:convergence2}, MILP shows superior convergence and optimization performance compared with Cuckoo Search, achieving lower objective function values and faster convergence.

\begin{figure}[!t]
        \centering
        \includegraphics[width = 0.3\textwidth, trim=0cm 0cm 0cm 0cm,clip]{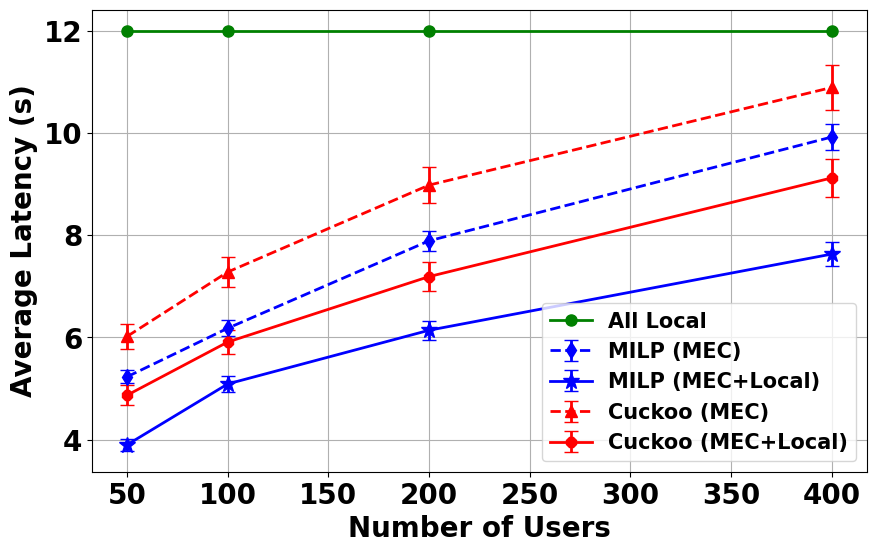}
        \caption{ Average latency based on number of users }
        \label{fig:latency} 
\end{figure}

\begin{figure}[!t]
        \centering
        \includegraphics[width = 0.3\textwidth, trim=0cm 0cm 0cm 0cm,clip]{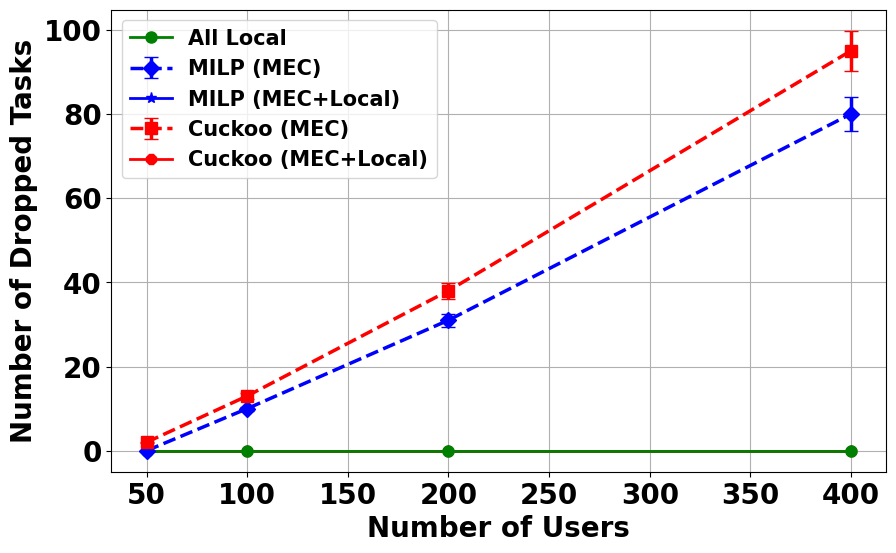}
        \caption{ Average number of dropped tasks based on number of users}
        \label{fig:drop} 
\end{figure}

 \figurename\hspace{0.1pt}\ref{fig:latency} illustrates the average latency for different number of users for 10 runs, comparing the performance of MILP and Cuckoo Search in two scenarios: MEC-only (offloading) and MEC+Local (partitioning) processing. As the number of users increase, the average latency continuously rises across every configuration, indicating the effects of increased system load. MILP frequently demonstrates less latency than Cuckoo Search, indicating its efficiency in task management and offloading. The latency of Cuckoo Search with MEC+Local processing is comparable to that of MILP in the MEC-only setup, especially for 50 and 100 users, indicating that Cuckoo Search can produce competitive outcomes when utilizing local resources. This discovery underscores that, with a suitable algorithm customized for the particular context, in our case MILP, offloading can achieve a low latency, close to that of partitioning.  The findings indicate that MILP is superior in reducing latency and managing increased user loads for both scenarios, especially in partitioning.
Additionally, the average number of dropped tasks for different number of users for both scenarios is plotted in \figurename\hspace{0.1pt}\ref{fig:drop}.  As the number of users increases, the number of dropped tasks rises significantly for both MILP and Cuckoo search when only MEC resources are utilized (offloading). However, MILP consistently shows fewer dropped tasks compared to Cuckoo Search in this configuration, demonstrating its superior capability in managing task allocation efficiently. In contrast, when local resources are incorporated along MEC servers (partitioning), both algorithms don't have any dropped task across all number of users. This highlights the effectiveness of leveraging local processing in our scenario to handle increased task loads and maintain service reliability. 

\begin{figure}[!hbt]
        \centering
        \includegraphics[width = 0.3\textwidth, trim=0cm 0cm 0cm 0cm,clip]{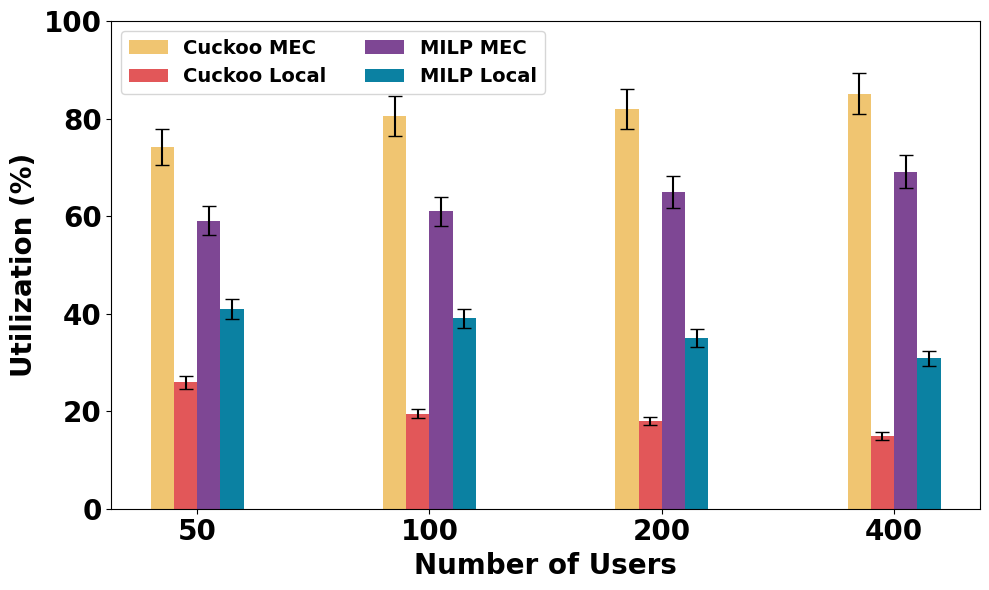}
        \caption{ Average MEC and local utilization in task partitioning }
        \label{fig:mec} 
\end{figure}

To understand how each algorithm manages the allocation of resources and distributes subtasks, we plot the utilization of both MEC and local resources in \figurename\hspace{0.1pt}\ref{fig:mec}.  The Cuckoo Search tends to rely predominantly on MEC resources, resulting in consistently high MEC utilization, while the local utilization remains low, suggesting less usage of local resources. On the other hand, MILP demonstrates a more balanced approach, showing high MEC utilization alongside significantly higher local utilization compared to Cuckoo Search. By leveraging local resources, MILP manages to reduce latency more than Cuckoo search and prevent task drops, showcasing its effectiveness in task management and resource usage.

\section{Conclusion}
The key challenges with task offloading in 5G-enabled MEC involve efficiently managing the need for lower latency and minimizing the the number of dropped tasks to ensure optimal system performance. Traditional full offloading or local processing approaches struggle to meet these requirements due to server overloads and unfulfilled service demands. To address these limitations, we introduced a task partitioning approach that optimally balances local processing with offloading to MEC servers. Additionally, we ensure that each user’s allocation of RBs remain within the available maximum number of RBs, while maintaining low latency. Our results showed that MILP partitioning achieved a 24\% latency reduction relative to full MILP offloading for high user loads, while Cuckoo partitioning delivered an 18\% latency improvement over Cuckoo offloading. These results affirm the effectiveness of our approach in optimizing task assignment and reducing system latency across varying number of users while ensuring no task drops. Looking forward, we intend to expand our research by incorporating energy consumption optimization alongside latency to further enhance the overall system efficiency and analyze the trade off between energy and latency for task partitioning when we ensure no task drops. Moreover, we may consider using machine learning to assist the model in offloading decision.


\section*{Acknowledgment}

This work was supported in part by funding from the Innovation for Defence Excellence and Security (IDEaS) program from the Department of National Defence (DND) and in part by Natural Sciences and Engineering Research Council of Canada (NSERC) CREATE TRAVERSAL Program.

\bibliographystyle{IEEEtran}

\end{document}